\documentstyle[a4,
twoside]{article}
\makeatletter
\renewcommand{\@oddhead}{ Notes on Rank One Perturbed Resolvent 
 \hfill \thepage}
\renewcommand{\@evenhead}{\thepage \hfill S. A. Choro\v{s}avin }
\renewcommand{\@oddfoot}{}
\renewcommand{\@evenfoot}{}
\makeatother

\author{S.A.~Choro\v{s}avin}
\title{ 
 Notes on Rank One Perturbed Resolvent.
 Perturbation of Isolated Eigenvalue.
 }
\date{}
\begin{document}
\maketitle 
\begin{abstract}
 This paper is a didactic commentary (a transcription with variations) 
 to the paper of S.R. Foguel 
 {\it Finite Dimensional Perturbations in Banach Spaces}. 

 Addressed, mainly: postgraduates and related readers.

 Subject:
 Suppose we have two linear operators,
$A, B$,
 so that 
$$
 B - A \mbox{ is rank one. }
$$
 Let 
$\lambda_o$
 be an {\it isolated} point of the spectrum of 
$A$:
$$
 \lambda_o \in \sigma(A) .
$$
 In addition, let 
$\lambda_o$
 be an {\it eigenvalue} of 
$A$:
$$
 \lambda_o \in \sigma_{pp}(A) .
$$

 The question is:
 Is 
$\lambda_o$
 in
$\sigma_{pp}(B)$ ? -- i.e., is 
$\lambda_o$
 an eigenvalue of 
$B$ ?

 And, if so, is the multiplicity of 
$\lambda_o$
 in 
$\sigma_{pp}(B)$
 equal to the multiplicity of 
$\lambda_o$
 in 
$\sigma_{pp}(A)$ ?
 -- or less?
 -- or greater?

\end{abstract}

\newpage
\section*%
{ Introduction }
 We continue to discuss the paper of 
 {\sc S.R. Foguel}, 
 {\it Finite Dimensional Perturbations in Banach Spaces},
 and we assume that the reader is familiar with our previous paper 
 arXiv:math-ph/0312016.

 The situatian we will discuss is:

\par\addvspace{2\bigskipamount}\par\noindent
 Let 
$A$
 and 
$B$,
 so that 
$A-B$
 is rank one
\footnote{ more accurately expressed, rank one or less },
 i.e.,
$$
 B-A = -f_a<l_a|
$$
 for an element 
$f_a$
 and a linear functional 
$l_a$.
 Next, let 
$\lambda_o$
 be an {\it isolated} point of the spectrum of 
$A$:
$$
 \lambda_o \in \sigma(A) .
$$
 In addition, let 
$\lambda_o$
 be an {\it eigenvalue} of 
$A$:
$$
 \lambda_o \in \sigma_{pp}(A) .
$$

\par\addvspace{2\bigskipamount}\par\noindent
 The question is:

\par\addvspace{2\bigskipamount}\par\noindent
 Is 
$\lambda_o$
 in
$\sigma_{pp}(B)$ ? -- i.e., is   
$\lambda_o$
 an eigenvalue of 
$B$ ?

\par\addvspace{\bigskipamount}\par\noindent
 And, if so, is the multiplicity of 
$\lambda_o$
 in 
$\sigma_{pp}(B)$
 equal to the multiplicity of 
$\lambda_o$
 in 
$\sigma_{pp}(A)$ ?
 -- or less?
 -- or greater?

\par\addvspace{2\bigskipamount}\par\noindent
 Foguel gave an answer in a very general situation. 
 We will not discuss all his constructions. 
 Instead, for technical reasons, we assume the underlying space 
$\cal H$
 to be {\it Hilbert}, and 
$A\,,\,B$
 to be bounded and symmetric, hence {\it self-adjoint},
 with respect to 
\begin{eqnarray*}
 (,) & = & \mbox{ Hilbert inner product on } {\cal H} .
\end{eqnarray*}

\par\addvspace{\medskipamount}\par\noindent
 Thus we restrict ourselves by discussing the situation 
 where there are fewer complications.

\newpage 

 Recall some facts.

\par\addvspace{2\bigskipamount}\par\noindent
 If 
${\cal A}$
 is a self-adjoint operator, and if 
$\lambda_o$
 is a {\bf non-real} number,
$Im\,\lambda_o \not= 0$, 
 then 
$$
 \lambda_o-{\cal A} : D({\cal A})\subset {\cal H} \to {\cal H}
$$
 is a bijection, and, in addition 
$$
 (\lambda_o-{\cal A})^{-1} : {\cal H} \to D({\cal A})\subset {\cal H}
$$
 is bounded.

\par\addvspace{2\bigskipamount}\par\noindent
 If 
${\cal A}$
 is a self-adjoint operator, and if 
$\lambda_o$
 is a {\bf real} number,
$Im\,\lambda_o = 0$, 
 then 
\begin{eqnarray*}
 P^{{\cal A}}_{\lambda_o}
 &:=&
 strong-\lim_{\epsilon\downarrow 0 }
 (i\epsilon(\lambda_o + i\epsilon-{\cal A})^{-1})
\end{eqnarray*}
 exists and the {\bf range} of 
$P^{{\cal A}}_{\lambda_o}$
 is the set of all 
$f_{\lambda_o}\in D({\cal A})$
 such that 
$$
 (\lambda_o -{\cal A})f_{\lambda_o} = 0 \,.
$$
 In addition, 
$P^{{\cal A}}_{\lambda_o}$
 is a {\it projection} operator and a {\it self-adjoint} operator.

\par\addvspace{2\bigskipamount}\par\noindent
 If 
$\lambda_o$
 is a {\bf real} number,
$Im\,\lambda_o = 0$, 
 so that for all 
$\lambda$
 near 
$\lambda_o$,
 and 
$\lambda \not= \lambda_o$,
 it has occurred that 
\begin{eqnarray*}
 \lambda \in \rho({\cal A})
 &=&
 \mbox{ the resolvent set of } {\cal A}\,,
\end{eqnarray*}
 in other words, if 
$ \lambda_o \in \rho({\cal A}) $
 or 
$ \lambda_o $
 is an {\it isolated } point of 
$\sigma({\cal A}) = \mbox{ spectrum of } {\cal A}$ ,
 then 
\begin{eqnarray*}
 (\lambda - {\cal A})^{-1}
 &=&
 \frac{P^{{\cal A}}_{\lambda_o}}{\lambda - \lambda_o}
\\&&{}
 + {\cal A}_{\lambda_o}
 - (\lambda - \lambda_o){\cal A}_{\lambda_o}^2
 + (\lambda - \lambda_o)^2{\cal A}_{\lambda_o}^3
 - \cdots
\\&&{}
 \mbox{ for all } \lambda \mbox{ near } \lambda_o \,,
 \lambda \not= \lambda_o \,,
\\&&{}
 \mbox{and for a {\bf bounded self-adjoint} } {\cal A}_{\lambda_o} \,.
\end{eqnarray*}
 In particular, if 
$\lambda_o$
 is an {\it isolated } point of 
$\sigma({\cal A})$ ,
 then 
$\lambda_o \in \sigma_{pp}({\cal A})$ .

\newpage

\par\addvspace{2\bigskipamount}\par\noindent
 Recall in addition, that

\par\addvspace{2\bigskipamount}\par\noindent
%
\parbox{\textwidth}{

\medskip

 if 
$A^{-1}$
 exists and 
$ 1-<l_a|A^{-1}f_a> \not= 0 $,
 then  
$B^{-1}$
 exists and, in addition,

\medskip

$$
 B^{-1}-A^{-1} = \frac{A^{-1}f_a<l_a|A^{-1}}{1-<l_a|A^{-1}f_a>}
$$ 
} 
%

\par\addvspace{2\bigskipamount}\par\noindent
 On the other hand, 

\medskip\noindent
\parbox{\textwidth}{

\medskip

$$
 \mbox{ if }
 1-<l_a|A^{-1}f_a> = 0 \,,
 \mbox{ then }
 B A^{-1}f_a =0 
$$
} 
%

\medskip\noindent
\parbox{\textwidth}{

\medskip

 \mbox{ if }
$$
 Bv_0 = 0
 \mbox{ and }
 v_0 \not= 0 \,,
$$
 \mbox{ then }
$$
 <l_a|v_0>\not=0 \,,\, 1-<l_a|A^{-1}f_a> =0 \,,\, BA^{-1}f_a =0\,,
 \mbox{ and }
 v_0 = A^{-1}f_a<l_a|v_0> \,.
$$
} 
%

\par\addvspace{2\bigskipamount}

 Notice that 
$$
 (\lambda-B)-(\lambda-A) = -(B-A) = f_a<l_a| \,.
$$
 Thus, we conclude:

\par\addvspace{2\bigskipamount}\par\noindent
\fbox{
\parbox{\textwidth}{

\medskip

 if 
$(\lambda-A)^{-1}$
 exists and 
$ 1+<l_a|(\lambda-A)^{-1}f_a> \not= 0 $,
 then  
$(\lambda-B)^{-1}$
 exists and, in addition,

\medskip

$$
 (\lambda-B)^{-1}-(\lambda-A)^{-1}
 =
 -\frac{(\lambda-A)^{-1}f_a<l_a|(\lambda-A)^{-1}}{1+<l_a|(\lambda-A)^{-1}f_a>}
$$ 
} 
} 

\par\addvspace{2\bigskipamount}\par\noindent
 On the other hand, 

\medskip\noindent
\fbox{
\parbox{\textwidth}{

\medskip

$$
 \mbox{ if }
 1+<l_a|(\lambda-A)^{-1}f_a> = 0 \,,
 \mbox{ then }
 (\lambda-B)(\lambda- A)^{-1}f_a =0 
$$
} 
} 

\medskip\noindent
\fbox{
\parbox{\textwidth}{

\medskip

 \mbox{ if }
$$
 (\lambda-B)v_0 = 0
 \mbox{ and }
 v_0 \not= 0 \,,
$$
 \mbox{ then }
$$
 <l_a|v_0>\not=0 \,,\, 1+<l_a|(\lambda-A)^{-1}f_a> =0
 \,,\, (\lambda-B)(\lambda-A)^{-1}f_a =0\,,
$$
 \mbox{ and }
$$
 v_0 = (\lambda-A)^{-1}f_a<l_a|v_0> \,.
$$
} 
} 
\newpage

\addvspace{2\bigskipamount}\par\noindent

 Before starting,
 we shall recall that we prefer Dirac's 
 "bra-ket" style of expressing, in the following form:

\par\addvspace{\bigskipamount}\par\noindent
{\it Notation 1}. \quad 
 If 
$f$
 is an element of a linear space, 
$X$,
 over a field, 
$K$, 
 then 
$|f>$
 stands for the mapping
$K \to X$,
 defined by 
$$
 |f>\lambda := \lambda f \quad .
$$

\par\addvspace{\bigskipamount}\par\noindent
{\it Notation 2}. \quad 
 If 
$l$
 is a functional 
 and we wish to emphasise this factor, then we write 
$<l|$ 
 instead of 
$l$.
 We also write 
$<l|f>$
 instead of 
$<l||f>1$,
 and write the terms 
$|f><l|$ 
 and 
$f<l|$
 interchangeably: 
$$
 <l|f> \equiv <l||f>1 \equiv l(f) \quad , \quad f<l| \equiv |f><l| \quad .
$$

\par\addvspace{\bigskipamount}\par\noindent
 Finally, we will restrict ourselves to the case where 
$<l_a|$
 is of the form:
\begin{eqnarray*}
 <l_a|u> &:=& \alpha(f_a|u) \quad ( u\in{\cal H} ) \,,\,
 \mbox{ for a real number } \alpha \,.
\end{eqnarray*}
 In this case, it is naturally to use the notations:
$$
 \alpha(f_a| :=<l_a| \mbox{ and } |f_a> := |f_a) \,. 
$$

\newpage 
\section
{\bf Perturbation of Isolated Eigenvalue. }

 Now, we turn to the relations, which links 
$(\lambda-A)^{-1}$
 and 
$(\lambda-B)^{-1}$,
 and which we now write 
 as follows:

\par\addvspace{\bigskipamount}\par\noindent
\fbox{
\parbox{\textwidth}{

\medskip

 if 
$(\lambda-A)^{-1}$
 exists and 
$ 1+\alpha(f_a|(\lambda-A)^{-1}f_a) \not= 0 $,
 then  
$(\lambda-B)^{-1}$
 exists and, in addition,

\medskip

$$
 (\lambda-\lambda_o)(\lambda-B)^{-1}-(\lambda-\lambda_o)(\lambda-A)^{-1}
 =
 -\alpha
 \frac{(\lambda-\lambda_o)(\lambda-A)^{-1}f_a
  (f_a|(\lambda-\lambda_o)(\lambda-A)^{-1}}{
 (\lambda-\lambda_o)\Bigl(1+\alpha(f_a|(\lambda-A)^{-1}f_a)\Bigr)}
$$ 
} 
} 

\par\addvspace{\bigskipamount}\par\noindent
 Note that the denominator is equal to 
%
\begin{eqnarray*}
 (\lambda-\lambda_o)
 & + &
 \alpha(f_a|P^{A}_{\lambda_o}f_a)
\\
 & + &
 (\lambda-\lambda_o) \alpha(f_a|{A}_{\lambda_o}f_a)
\\
 & - &
 (\lambda-\lambda_o)^2 \alpha(f_a|{A}_{\lambda_o}^2f_a)
 + (\lambda-\lambda_o)^3 \alpha(f_a|{A}_{\lambda_o}^3f_a)
 - \cdots
\end{eqnarray*}

\par\addvspace{2\bigskipamount}\par\noindent

 We distinguish three cases:

\par\addvspace{2\bigskipamount}\par\noindent
 (a) 
$$
 \alpha(f_a|P^{A}_{\lambda_o}f_a) \not= 0
$$

\par\addvspace{\bigskipamount}\par\noindent
 (b) 
$$
 \alpha(f_a|P^{A}_{\lambda_o}f_a) = 0 \,,
 1 +  \alpha(f_a|{A}_{\lambda_o}f_a) \not= 0
$$

\par\addvspace{\bigskipamount}\par\noindent
 (c) 
$$
 (f_a|P^{A}_{\lambda_o}f_a) = 0 \,,\,
 1 +  \alpha(f_a|{A}_{\lambda_o}f_a) = 0 \,,\,
 \alpha \not= 0 \,.
$$

\par\addvspace{\bigskipamount}\par\noindent
 It is worthy to note that if 
$$
 1 +  \alpha(f_a|{A}_{\lambda_o}f_a) = 0 \,
$$
 then 
$$
 1 =|\alpha(f_a|{A}_{\lambda_o}f_a)|^2
 \leq |\alpha|^2 (f_a|f_a)({A}_{\lambda_o}f_a|{A}_{\lambda_o}f_a)
 = |\alpha|^2 (f_a|f_a)(f_a|{A}_{\lambda_o}^2f_a)
$$
 As a result,
$$
 \mbox{ if }
 1 +  \alpha(f_a|{A}_{\lambda_o}f_a) = 0 \,,
 \mbox{ then }
 (f_a|{A}_{\lambda_o}^2f_a) \not=0
$$

\par\addvspace{2\bigskipamount}\par\noindent
 Now let 
$$
 \lambda := \lambda_o + i\epsilon \mbox{ and }  \epsilon\downarrow 0 \,.
$$

\par\addvspace{2\bigskipamount}\par\noindent
 Then we infer:

\newpage 
%
 (a) 
$$
 \alpha(f_a|P^{A}_{\lambda_o}f_a) \not= 0
$$
 In this case,
\begin{eqnarray*}
 P^{B}_{\lambda_o}-P^{A}_{\lambda_o}
 &=&
 -\alpha
 \frac{P^{A}_{\lambda_o}f_a
  (f_a|P^{A}_{\lambda_o}}{\alpha(f_a|P^{A}_{\lambda_o}f_a)}
\\
 &=&
 -
 \frac{P^{A}_{\lambda_o}f_a
  (f_a|P^{A}_{\lambda_o}}{(f_a|P^{A}_{\lambda_o}f_a)}
 \quad (\mbox{note that } \alpha \not= 0 )
\end{eqnarray*} 
 In particular, 
$$
 dim P^{B}_{\lambda_o} = dim P^{A}_{\lambda_o} -1 \,.
$$

\par\addvspace{2\bigskipamount}\par\noindent
 (b) 
$$
 \alpha(f_a|P^{A}_{\lambda_o}f_a) = 0 \,,\,
 1 +  \alpha(f_a|{A}_{\lambda_o}f_a) \not= 0 \,.
$$
 In this case, if 
$\alpha =0$,
 then 
$ B=A $, 
$ P^{B}_{\lambda_o} = P^{A}_{\lambda_o} $,
$ dim P^{B}_{\lambda_o} = dim P^{A}_{\lambda_o} $.
 Otherwise,
$(f_a|P^{A}_{\lambda_o}f_a) = 0$
 and 
$$
 (P^{A}_{\lambda_o}f_a|P^{A}_{\lambda_o}f_a)
 = 
 (f_a|P^{A}_{\lambda_o}f_a) = 0 \,.
$$
 Hence 
$$
 P^{A}_{\lambda_o}f_a = 0 \,,
$$
 and 
\begin{eqnarray*}
 (\lambda - A)^{-1}f_a
 &=&
 + A_{\lambda_o}f_a
 - (\lambda - \lambda_o)A_{\lambda_o}^2f_a
 + (\lambda - \lambda_o)^2A_{\lambda_o}^3f_a
 - \cdots
\\&&{}
 \mbox{ for all } \lambda \mbox{ near } \lambda_o \,,
 \lambda \not= \lambda_o \,,
\\&&{}
 \mbox{and for a {\bf bounded self-adjoint} } A_{\lambda_o} \,.
\end{eqnarray*}
\begin{eqnarray*}
 P^{B}_{\lambda_o}-P^{A}_{\lambda_o}
 &=&
 0 \,.
\end{eqnarray*} 
 In particular, 
$$
 dim P^{B}_{\lambda_o} = dim P^{A}_{\lambda_o} \,,
$$
 as well.

\par\addvspace{2\bigskipamount}\par\noindent
 (c) 
$$
 (f_a|P^{A}_{\lambda_o}f_a) = 0 \,,\,
 1 +  \alpha(f_a|{A}_{\lambda_o}f_a) = 0 \,,\,
 \alpha \not= 0 \,.
$$

 In this case, as well as in the case (b), 
$$
 P^{A}_{\lambda_o}f_a = 0 \,,
$$
 and 
\begin{eqnarray*}
 (\lambda - A)^{-1}f_a
 &=&
 + A_{\lambda_o}f_a
 - (\lambda - \lambda_o)A_{\lambda_o}^2f_a
 + (\lambda - \lambda_o)^2A_{\lambda_o}^3f_a
 - \cdots
\\&&{}
 \mbox{ for all } \lambda \mbox{ near } \lambda_o \,,
 \lambda \not= \lambda_o \,,
\end{eqnarray*}
 However 
\begin{eqnarray*}
 P^{B}_{\lambda_o}-P^{A}_{\lambda_o}
 &=&
 -\alpha
 \frac{A_{\lambda_o}f_a
  (f_a|A_{\lambda_o}}{-\alpha(f_a|A_{\lambda_o}^2f_a)}
\\
 &=&
 \frac{A_{\lambda_o}f_a
  (f_a|A_{\lambda_o}}{(f_a|A_{\lambda_o}^2f_a)}
 \quad (\alpha \not= 0 )
\end{eqnarray*} 
 In particular, 
$$
 dim P^{B}_{\lambda_o} = dim P^{A}_{\lambda_o} +1 \,.
$$

\par\addvspace{2\bigskipamount}\par\noindent

\newpage 
\section
{\bf Example }

\addvspace{\medskipamount}\par\noindent
 Let
${\cal T}$
 stands for the functions transformation defined by 
$$
  ({\cal T}u)(x) := -\frac{\partial^2 u(x)}{\partial x^2 } \,;
$$
${\cal T}_{DD}$
 be the restriction of 
${\cal T}$
 so that 
${\cal T}_{DD}$
 acts on that functions,
$u$,
 for which
$({\cal T}u)(x)$
 is defined at 
$ 0 \leq x \leq 1 $
 and, in addition: 
\begin{eqnarray*}
 u(0) &=& 0
\\
 u(1) &=& 0
\end{eqnarray*}
 We take
$L_2(0,1)$, as the underlying space 
${\cal H}$.
 In this space, 
${\cal T}_{DD}$
 is closable and symmetric.
 Moreover, it is essentially self-adjoint. 
 That means that its closure,
$T_{DD}$,
 is self-adjoint.

 One can check that 
$ T_{DD}^{-1} $
 exists and 
 is an integral operator; its integral kernel is 
\begin{eqnarray*}
 G_{DD}(x,\xi)
 &=&
 -
   \left\{\begin{array}{rcl}
   x\cdot (\xi-1) &,& \mbox{ if }   x \leq \xi \\
   (x-1)\cdot \xi &,& \mbox{ if } \xi \leq x \\
   \end{array}\right\}
\end{eqnarray*}
 One can also check that 
$$
 \sin(\pi x), \sin(2\pi x),\sin(3\pi x), \ldots
$$
 are eigenfunctions of 
$T_{DD}$
 and, of course, of 
$T_{DD}^{-1}$.
 The corresponding eigenvalues of 
$T_{DD}$
 are 
$$
 \pi^2 , (2\pi)^2, (3\pi)^2, \ldots
$$
 and that of 
$T_{DD}^{-1}$
 are 
$$
 \frac{1}{\pi^2} , \frac{1}{(2\pi)^2}, \frac{1}{(3\pi)^2}, \ldots
$$
 All eigenvalues are multiplicity-free.
 It is not very difficult to describe 
$$
 (z - T_{DD})^{-1} \,.
$$ 
 This is an integral operator. Its integral kernel is 
\begin{eqnarray*}
 G_{DD}(x,\xi,z)
 &=&
 -\frac{1}{k\sin(k)}
   \left\{\begin{array}{rcl}
   \sin(kx)\sin(k(1-\xi)) &,& \mbox{ if }   x \leq \xi \\
   \sin(k\xi)\sin(k(1-x))  &,& \mbox{ if } \xi \leq x \\
   \end{array}\right\}
 \\&&{} 
 \mbox{ where $k$ is defined by } k^2 = z \,,
\end{eqnarray*}
 and where, of course, 
$z$
 is to be so, that 
$$
 \sin(k) \not= 0 \,.
$$
 As for  
$$
 (\lambda  -T_{DD}^{-1})^{-1} \,,
$$ 
 it is not very difficult to describe it as well: 
 A general (and quite standard) argumentation is:
\begin{eqnarray*}
 (\lambda -T_{DD}^{-1})^{-1}
&=&
 T_{DD}(\lambda T_{DD}-I)^{-1}
\\&&{}=
\frac{1}{\lambda }\Bigl(
 \lambda T_{DD}(\lambda T_{DD}-I)^{-1}
\Bigr)
\\&&{}=
\frac{1}{\lambda }\Bigl(
 (\lambda T_{DD}-I+I)(\lambda T_{DD}-I)^{-1}
\Bigr)
\\&&{}=
\frac{1}{\lambda }\Bigl(
 I+(\lambda T_{DD}-I)^{-1}
\Bigr)
\\&&{}=
\frac{1}{\lambda }\Bigl(
 I-\frac{1}{\lambda }(\frac{1}{\lambda }-T_{DD})^{-1}
\Bigr)
 \quad (\mbox{ naturally, here } \lambda  \not= 0 ) \,.
\end{eqnarray*}

\newpage
 Now, we let 
\begin{eqnarray*}
 A &:=& T_{DD}^{-1} \\
 B &:=& A +\alpha f_a(f_a|
\end{eqnarray*}
 where 
$f_a$
 is defined by 
$$
 f_a(x):= (x-\frac{1}{2})  \,,
$$
 and let us apply the theory described in the previous section. So, let 
$$
 f_z := z(z-T_{DD})^{-1}f_a \,,
$$
 i.e.,
\begin{eqnarray*}
 zf_z(x) + \frac{\partial^2 f_z(x)}{\partial x^2}
&=&
 z(x-\frac{1}{2}) \,,
\\
 f_z(0) 
&=&
 0 \,,
\\
 f_z(1) 
&=&
 0 \,.
\end{eqnarray*}
 Hence 
\begin{eqnarray*}
 f_z(x)
&=&
 (x-\frac{1}{2})
 -\frac{1}{2}\frac{\sin(k(x-\frac{1}{2}))}{\sin(k(\frac{1}{2}))}
\\&&{}
 \mbox{ where $k$ is defined by } k^2 = z \,,
\end{eqnarray*}
 and where, recall, 
$z$
 is such that 
$$
 \sin(k) \not= 0 \,.
$$
 Thus we deduce:
\begin{eqnarray*}
 \Bigl((-I+z(z-T_{DD})^{-1})f_a\Bigr)(x)
&=&
 -f(x) +f_z(x)
\\
&=&
 -(x-\frac{1}{2})
 +(x-\frac{1}{2})
 -\frac{1}{2}\frac{\sin(k(x-\frac{1}{2}))}{\sin(k(\frac{1}{2}))}
\\
&=&
 -\frac{1}{2}\frac{\sin(k(x-\frac{1}{2}))}{\sin(k(\frac{1}{2}))}
\\&&{}
 \mbox{ where $k$ is defined by } k^2 = z \,,
\end{eqnarray*}
\begin{eqnarray*}
 (f_a|(-I+z(z-T_{DD})^{-1})f_a)
&=&
 -\frac{1}{2}\int_0^1 (\xi-\frac{1}{2})
 \frac{\sin(k(\xi-\frac{1}{2}))}{\sin(k(\frac{1}{2}))} d\xi
\\
&=&
 \frac{1}{2}\int_{\xi=0}^1 (\xi-\frac{1}{2})
 \frac{d\cos(k(\xi-\frac{1}{2}))}{k\sin(k(\frac{1}{2}))}
\\
&=&
 \frac{1}{2}
 \frac{\cos(k(\frac{1}{2}))}{k\sin(k(\frac{1}{2}))}
 - \frac{1}{2}
 \int_{\xi=0}^1 \frac{\cos(k(\xi-\frac{1}{2}))}{k\sin(k(\frac{1}{2}))}
 d(\xi-\frac{1}{2})
\\
&=&
 \frac{1}{2}
 \frac{\frac{1}{2}\cos(k(\frac{1}{2}))}{k\sin(k(\frac{1}{2}))}
 - \frac{1}{k^2}
\end{eqnarray*}
\begin{eqnarray*}
 1+z\alpha(f_a|(-I+z(z-T_{DD})^{-1})f_a)
&=&
 1+k^2\alpha\Bigl(
 \frac{\frac{1}{2}\cos(k(\frac{1}{2}))}{k\sin(k(\frac{1}{2}))}
 -\frac{1}{k^2}\Bigr)
\\
&=&
 1+\alpha\Bigl(
 \frac{k}{2}\frac{\cos(\frac{k}{2})}{\sin(\frac{k}{2})}
 -1 \Bigr)
\end{eqnarray*}

\addvspace{\medskipamount}\par\noindent

 {\it We conclude} :

\addvspace{\medskipamount}\par\noindent
 The new eigenvalues,
$\lambda_n$,
 are defined by 
$$
 1+z_n\alpha (f_a|(-I+z_n(z_n-T_{DD})^{-1})f_a) =0 \,,
$$
 i.e., by 
$$
 1+\alpha\Bigl(
 \frac{k_n}{2}\frac{\cos(\frac{k_n}{2})}{\sin(\frac{k_n}{2})}
 -1 \Bigr)
 = 0 \,,
$$
 and the associated eigenfunctions are 
\begin{eqnarray*}
 \Bigl((-I+z_n(z_n-T_{DD})^{-1})f_a\Bigr)(x)
&=&
 -\frac{1}{2}\frac{\sin(k_n(x-\frac{1}{2}))}{\sin(k_n(\frac{1}{2}))}
\end{eqnarray*}

\par\addvspace{\medskipamount}\par\noindent
 Now, let 
\begin{eqnarray*}
 \alpha &:=& 1 \,.
\end{eqnarray*}
 In this case, the solutions to 
$$
 1+\alpha\Bigl(
 \frac{k_n}{2}\frac{\cos(\frac{k_n}{2})}{\sin(\frac{k_n}{2})}
 -1 \Bigr)
 = 0
$$
 are 
\begin{eqnarray*}
 k_n &=& \pi, 3\pi, 5\pi, \ldots \,,
\end{eqnarray*}
 which conflicts with 
\begin{eqnarray*}
 \lambda_n & \in & \rho(T_{DD}^{-1}) \,.
\end{eqnarray*}
 Hence, there are {\bf no new eigenvalues}, if 
$\alpha = 1$\quad (!!) 

\par\addvspace{\medskipamount}\par\noindent
 Now, let us analyse
$$
 \lambda_o - B
$$
 in the case where 
\begin{eqnarray*}
 \lambda_o & \in & \sigma(T_{DD}^{-1}) \,,  \lambda_o \not= 0 \,,
\end{eqnarray*}
 i.e., where 
\begin{eqnarray*}
 k_o & \in & \{ \pi, 2\pi, 3\pi, 4\pi \ldots \} \,.
\end{eqnarray*}

\par\addvspace{2\bigskipamount}\par\noindent
 Firstly we notice that 
$\lambda_o$
 is multiplicity-free (in 
$\sigma(A)$)
 and that a corresponding eigenfunction,
$f_{\lambda_o}$,
 is such that 
\begin{eqnarray*} 
 f_{\lambda_o}(x)
 & = &
 \sin(k_o x) \quad (x \in [0,1]) \,.
\end{eqnarray*} 
 We have:
$(f_{\lambda_o}|f_{\lambda_o})=1/2$,
 hence,
\begin{eqnarray*} 
 P^{{\cal A}}_{\lambda_o}
 & = &
 2f_{\lambda_o}(f_{\lambda_o}|
\end{eqnarray*} 

\newpage

 In accordance with the scheme which had been formed 
 in the previous section, we analyse 
$ (f_a|P^{A}_{\lambda_o}f_a) $ .
 We have, 
\begin{eqnarray*} 
 (f_a|P^{A}_{\lambda_o}f_a) 
 & = &
 (f_a|2f_{\lambda_o})(f_{\lambda_o}|f_a)
\\&&{}
 =2\Bigl|\int_0^1(\xi-\frac{1}{2})\sin(k_o\xi) d\xi\Bigr|^2
\\&&{}
 =2\Biggl|
  \Bigl(
 -\frac{\xi}{k_o}\cos(k_o\xi)
 +\frac{1}{k_o^2}\sin(k_o\xi)
 +\frac{1}{2k_o}\cos(k_o\xi)
 \Bigr)
 \Big|_{\xi=0}^{\xi=1}
 \Biggr|^2
\\&&{}
 =
 2\Bigl|
 \frac{1}{2k_o}\bigl(1+\cos(k_o)\bigr)
 \Bigr|^2
 \,.
\end{eqnarray*} 

\par\addvspace{\bigskipamount}\par\noindent
 We divide the analysis into two parts.
\par\addvspace{\bigskipamount}\par\noindent
 If 
\begin{eqnarray*}
 k_o & \in & \{ 2\pi, 4\pi, 6\pi \ldots \} \,,
\end{eqnarray*}
 then 
\begin{eqnarray*}
 (f_a|P^{A}_{\lambda_o}f_a) 
 & = &
 \frac{2}{k_o^2} \,.
\end{eqnarray*}
 In this case, 
\begin{eqnarray*} 
 (f_a|P^{A}_{\lambda_o}f_a) 
 & \not= &
 0 \,.
\end{eqnarray*} 
 Hence, if 
\begin{eqnarray*}
 k_o & \in & \{ 2\pi, 4\pi, 6\pi \ldots \} \,,
\end{eqnarray*}
 then 
$$
 dim P^{B}_{\lambda_o} = dim P^{A}_{\lambda_o} -1 = 1-1 = 0
$$
 and 
$$
 \lambda_o \mbox{ is no eigenvalue of } B \,.
$$

\par\addvspace{2\bigskipamount}\par\noindent
 On the contrary, if 
\begin{eqnarray*}
 k_o & \in & \{ \pi, 3\pi, 5\pi \ldots \} \,,
\end{eqnarray*}
 then 
\begin{eqnarray*} 
 (f_a|P^{A}_{\lambda_o}f_a) 
 & = &
 0 \,.
\end{eqnarray*} 
 Hence, if 
\begin{eqnarray*}
 k_o & \in & \{ \pi, 3\pi, 5\pi \ldots \} \,,
\end{eqnarray*}
 then 
$$
 dim P^{B}_{\lambda_o} \geq dim P^{A}_{\lambda_o} = 1
$$
 and 
$$
 \lambda_o \mbox{ is an eigenvalue of } B \,.
$$%

\par\addvspace{2\bigskipamount}\par\noindent
 And what is equal then the multiplicity of 
$\lambda_o$ to?
\par\addvspace{2\bigskipamount}\par\noindent
 To answer this question, let us analyse  
\begin{eqnarray*} 
 1 +  \alpha(f_a|{A}_{\lambda_o}f_a)
 & = &
 1 + \lim_{\epsilon\downarrow 0 }
 (f_a|
 \Bigl(
 (\lambda_o + i\epsilon - A)^{-1}
 -
 \frac{P^{{\cal A}}_{\lambda_o}}{i\epsilon}
 \Bigr)
 |f_a)
\\
 & = &
 1 + \lim_{\epsilon\downarrow 0 }
 (f_a|
 \Bigl(
 (\lambda_o + i\epsilon - A)^{-1}
 \Bigr)
 |f_a)
\end{eqnarray*} 
 Notice, 
\begin{eqnarray*}
 1 +
 (f_a|
 \Bigl(
 (\lambda - A)^{-1}
 \Bigr)
 |f_a)
&=&
 1+z\alpha(f_a|(-I+z(z-T_{DD})^{-1})f_a)
\\
&=&
 1+k^2\alpha\Bigl(
 \frac{\frac{1}{2}\cos(k(\frac{1}{2}))}{k\sin(k(\frac{1}{2}))}
 -\frac{1}{k^2}\Bigr)
\\
&=&
 1+\alpha\Bigl(
 \frac{k}{2}\frac{\cos(\frac{k}{2})}{\sin(\frac{k}{2})}
 -1 \Bigr)
\\
&=&
 \frac{k}{2}\frac{\cos(\frac{k}{2})}{\sin(\frac{k}{2})}
 \quad (\mbox{ because } \alpha =1 ) \,.
\end{eqnarray*}
 Hence, 
\begin{eqnarray*} 
 1 +  \alpha(f_a|{A}_{\lambda_o}f_a)
& = &
 0 \,.
\end{eqnarray*}
\par\addvspace{\bigskipamount}\par\noindent
 Thus, we conclude:

\par\addvspace{\bigskipamount}\par\noindent
 If
\begin{eqnarray*} 
 k_o & \in & \{ \pi, 3\pi, 5\pi \ldots \} \,,
\end{eqnarray*}
 then
\begin{eqnarray*} 
 \mbox{ the multiplicity of } \lambda_o &=& 2 \,.
\end{eqnarray*}

\par\addvspace{2\bigskipamount}\par\noindent
{\it
 Exercise. \quad Compare 
$B$
 with the Green's function generated by 
$-\frac{\partial^2}{\partial x^2}$
 on 
$[0,1]$
 and relations 
\begin{eqnarray*}
 u(0)=-u(1) &,&
 \frac{\partial u}{\partial x}\Big|_{x=0}=-
 \frac{\partial u}{\partial x}\Big|_{x=1} \,.
\end{eqnarray*}
}

\par\addvspace{2\bigskipamount}\par\noindent

\bibliographystyle{unsrt}

\end{document}